\documentclass[twocolumn,showpacs,prb]{revtex4}

\usepackage{graphicx}
\usepackage{dcolumn}
\usepackage{amsmath}

\begin{document}

\title{Weak Localization in an Ultradense 2D Electron Gas in $\delta$-doped Silicon}
\author{M. A. Zudov}
\author{C. L. Yang}
\author{R. R. Du}
\affiliation{Department of Physics, University of Utah, Salt Lake City, Utah 84112}
\author{T.-C. Shen}
\author{J.-Y. Ji}
\affiliation{Department of Physics, Utah State University, Logan, Utah 84322}
\author{J. S. Kline}
\author{J. R. Tucker}
\affiliation{Department of Electrical and Computer Engineering, University of Illinois, Urbana, Illinois 61801}

\received{May 16, 2003}

\begin{abstract}
An ultradense 2D electron system can be realized by adsorbing PH$_3$ precursor molecules onto an atomically clean Si surface, followed by epitaxial Si overgrowth.
By controlling the PH$_3$ coverage the carrier density of such system can easily reach $\sim$10$^{14}$ cm$^{-2}$, exceeding that typically found in GaAs/AlGaAs structures by more than two-three orders of magnitude.
We report on a first systematic characterization of such novel system by means of standard magnetotransport.
The main findings include logarithmic temperature dependence of zero-field conductivity and logarithmic negative magnetoresistance.
We analyzed the results in terms of scaling theory of localization in two dimensions.
\end{abstract}

\maketitle


Physics of 2D electron systems (2DES) have been traditionally explored using the cleanest systems available to date, most exclusively GaAs/AlGaAs-based heterostructures and quantum wells.\cite{rev}
While the mobility of GaAs/AlGaAs systems can be very high, in excess of $10^7$ cm$^2$/Vs, the carrier density is typically limited to the $10^{11}$ cm$^{-2}$ range.
During the last decade, 2D systems of much greater density have been fabricated in Si, e.g. by overgrowing boron ($p$-type)\cite{weir} or antimony ($n$-type)\cite{citrin} adlayers at low temperatures ($T\sim 450$ K).
Recently, we have reported on a fabrication of an ultradense Si-based 2DES in which the electrons are donated by a layer of phosphorous atoms.\cite{tcapl}
The ability of e-beam patterned Si hydride surfaces to selectively adsorb the PH$_3$ makes this particular system especially promising for fabrication of atom-scale electronic structures in single-crystal Si.\cite{ts}
In the fabrication process, a room temperature adsorption of PH$_3$ precursor molecules onto an atomically flat Si surface was employed to deposit the $\delta$-layer, followed by low-temperature overgrowth.
Since the carriers remain confined to a dopant plane by electrical fields, strong short-range scattering significantly limits the mobility (typically, $\mu \le 10^2$ cm$^2$/Vs).
On the other hand, this is a rather unusual system, as the carrier density, $n$, can easily reach $10^{14}$ cm$^{-2}$ exceeding those typically found in GaAs-based systems by two orders of magnitude.

In this paper we report on a first systematic electrical characterization of an ultradense 2DES residing on a $\delta$-layer of P in single-crystal Si.
Employing conventional magnetotransport methods, we have performed conductivity measurements in a variety of experimental conditions, at temperatures, $T$, from 0.3 to 100 K, and in magnetic fields of up to 5 T.
The temperature dependence of the zero-field conductivity showed a drastic reduction between 30 and 20 K in {\em all} samples followed by logarithmic behavior at lower $T$.
In addition, negative magnetoresistance was observed in magnetic fields up to 5 T which exhibited the characteristic logarithmic dependence as well.
While observed logarithmic regimes strongly suggest weak localization in such systems, there are some signatures of the importance of the correlation effects.
In particular, the enhancement of the normalized coefficient $p$ in the $\ln$($T$/K) correction may point out onto electron-electron interaction effects, which were previously thought to be unimportant.

\begin{figure}[b]
\includegraphics{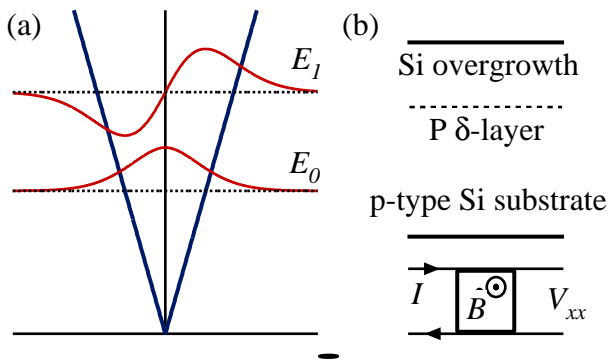}
\caption{(a) Confinement potential and electronic states in $\delta$-doped Si.
(b) Schematics of the sample structure (top) and magnetotransport measurements (bottom).
}
\label{fig0}
\end{figure}

Fig. \ref{fig0} depicts the confinement potential due to ionized $\delta$-layer and the electron wavefunctions in the lowest two subbands.
Parameters of our samples are presented in Table 1.
The dimensionless parameter, $r_s=[a_B^*\sqrt{\pi n}]^{-1}$, is the ratio of the Coulomb interaction energy to the Fermi energy, where $a_B^*=\epsilon\hbar^2/m^*e^2$ is the effective Bohr radius.
An effective mass $m^*=0.26m_0$ and a dielectric constant $\epsilon=11.9$ are used.
The smallness of $r_s$ ($< 1$) indicates that our 2DES is in the ultradense regime.
\begin{table}[bth]
\caption{PH$_3$ coverages, densities, $r_s$, and mobilities of the samples studied. The PH$_3$ coverages were measured by STM. The densities and mobilities were obtained by Hall measurements at $T\approx 4$ K. }
\label{table1}
\begin{ruledtabular}
\begin{tabular*}{\hsize}{l@{\extracolsep{0ptplus1fil}}c@{\extracolsep{
0ptplus1fil}}c@{\extracolsep{0ptplus1fil}}c@{\extracolsep{0ptplus1fil}}c}
Sample & A & B & C\\
\colrule
PH$_3$ Coverage, \% & 15 & 40 & 4\\
$n$, 10$^{14}$ cm$^{-2}$ & 0.88 & 1.31 & 0.36\\
$r_s$ & 0.25 & 0.20 & 0.39\\
$\mu$, cm$^2$/Vs & $\sim$38 & $\sim$28 & $\sim$13\\
\end{tabular*}
\end{ruledtabular}
\end{table}

By studying the correlation between the PH$_3$ coverage and the carrier density, obtained from Hall measurements, we have earlier concluded\cite{tcapl} that samples became insulating when average inter-particle spacing approaches the Bohr radius in Si ($r_{{\rm B}} \approx 25$ \AA), suggesting small redistribution of P atoms during Si overgrowth.
Sample C presented here has the lowest PH$_3$ coverage we have achieved to ensure the conduction at low $T$.
On the other hand, we have also noticed\cite{tcapl} that raising the PH$_3$ coverage above 30\% may result in density {\em reduction} due to formation of defects causing incomplete donor activation.
This situation is realized in sample B, where only about 50\% of donors are electrically active.

To make electrical connections to the ultradense 2DES we evaporated four AuSb pads along the perimeter of a square ($\sim$3 mm by 3 mm) sample and then annealed at 200 $^\circ$C for ten minutes.
Resistance as a function of temperature and magnetic field was measured in a $^3$He/14T superconducting magnet system using a conventional four-terminal van der Pauw technique(most appropriate for our sample geometry) and low-frequency (7-13 Hz) lock-in detection.
\begin{figure}[b]
\includegraphics{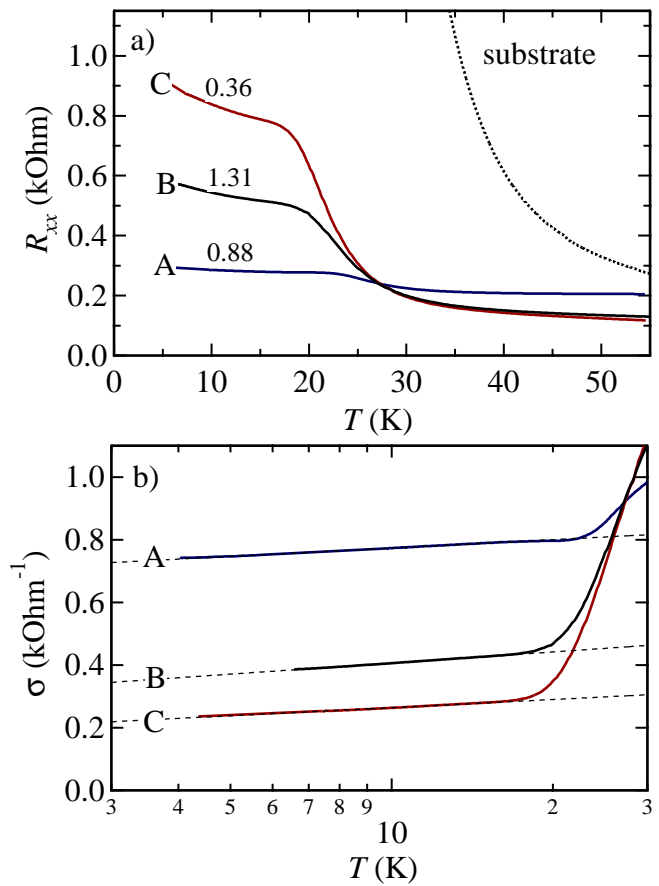}
\caption{(a) Zero-field resistance for samples A, B, C (solid lines), and Si substrate (dotted line) as a function of temperature.
The traces are labeled with electron density in units of $10^{14}$ cm$^{-2}$.
All samples show steep increase in $R_{xx}$ between 30 and 20 K, with a ``crossing point'' at $T^*\approx 27$ K, after which resistance tends to saturate and remains finite down to low temperatures; as expected, the substrate shows insulating behavior.
(b) Conductivity of samples A, B, and C plotted on a semi-log scale as function of temperature.
Below $\sim 20$ K all samples show logarithmic temperature dependence of the conductivity with roughly the same slope.
Dotted lines represent fits to the logarithmic regions using Eqn.~\ref{eqn1}.
}
\label{fig1}
\end{figure}

In Fig.~\ref{fig1}(a) we present zero-field resistance, $R_{xx}$, for samples A, B, and C, as a function of temperature.
For comparison we also include the data from the Si substrate shown by dotted line.
As expected, the substrate becomes insulating at low $T$, which confirms that carriers in other samples are from the phosphorous $\delta$-layer embedded into the Si crystal.
On the other hand, resistance of the samples (A, B, and C) show a tendency to saturate, after steep increase between 30 and 20 K, and remains finite down to low temperatures.
We also note that while samples have different parameters, the resistance traces cross in one point, $T^*\approx 27$ K.
In Fig.~\ref{fig1}(b) we present the conductivity, $\sigma_{xx}=1/\rho_{xx}$ ($\rho_{xx}=\pi R_{xx}$/\rm{ln}2), for samples A, B, and C, converted from the resistance data shown in Fig.~\ref{fig1}(a).
Plotted on a semi-log scale, data from all the samples under study show roughly linear dependence at $T < 20$ K.
Such logarithmic behavior has been routinely observed in thin metal films\cite{cu, pt} and was explained in terms of scaling theory of localization in two dimensions.\cite{lr}
We emphasize that the logarithmic $T$-dependence, that is observed in our samples below 20 K, is exclusive for 2D, further confirming two-dimentionality of our system.
As was pointed out by Thouless,\cite{th} inelastic scattering processes give rise to random fluctuations in the time evolution of an electronic state therefore suppressing quantum interference necessary for localization.
In this picture, assuming that elastic scattering dominates, e.g. $\tau \ll \tau_{{\rm in}}$, the electron diffuses a distance, called dephasing length, $L_{{\rm th}}=\sqrt{D\tau_{{\rm in}}}$, where $D$ is the diffusion constant.
Inelastic scattering time typically increases at low temperatures as $\tau_{{\rm in}} \propto T^{-p}$), an therefore the dephasing length becomes $T$-dependent, $L_{{\rm th}} \propto T^{-p/2}$.
As scale-dependent localization effects are limited to this lengthscale, the temperature dependence of the 2D conductivity is expected to be logarithmic:\cite{lr}
\begin{equation}
\sigma(T)=\sigma_0+\frac{pe^2}{\pi h}{\rm ln}\left(\frac{T}{T_0}\right)
\label{eqn1}
\end{equation}
where $\sigma_0=ne^2\tau/m^*$ is Drude conductivity ($n$ and $m^*$ are carrier density and effective mass respectively) and $p\sim 1$.
In addition to that, there is also exist a correction due to electron-electron correlation effects which is also logarithmic with $T$ in 2D \cite{lr}.
Since to the lowest order the corrections are additive, the parameter $p$ will be modified if electron-electron interactions are present.

Due to extremely high density of our system, one would normally expect non-interacting behavior, i.e. $p\sim 1$.
However, fitting our data with Eqn.~\ref{eqn1} (cf. dotted lines in Fig.~\ref{fig1}(b)) results in $p \sim 3-4$ for all samples (i.e., 3.1, 4.1, and 3.0 for A, B, and C, respectively).
The fact that $p$ for sample B is greater than in other samples may be related to the considerably higher donor coverage and/or incomplete donor activation.
Previous studies on Cu\cite{cu} and Pt\cite{pt} thin films, as well as on ultradense 2D hole system,\cite{weir} all reported $p\sim 1$.
To distinguish between the weak localization and electron-electron correlation effects, one typically has to perform magnetotransport studies, as the localization is also expected to produce characteristic negative magnetoresistance.

In a magnetic field a new length scale appears, namely the size of the first Landau orbit, or magnetic length, $\lambda=\sqrt{\hbar/eB}$, which decreases with magnetic field.
Once its size becomes comparable to the dephasing length, $L_{{\rm Th}}$ the localization is suppressed and the corresponding correction to the conductivity is described by:\cite{alt}
\begin{equation}
\delta \sigma(B)=\frac{e^2}{\pi h}\left[\psi\left(\frac{1}{2}+\frac{1}{x}\right)+{\rm ln}x\right]
\label{eqn2}
\end{equation}
where $\psi$ denotes the digamma function and $x=(2L_{{\rm Th}}/\lambda)^2\propto B$.
In the low field limit ($x\ll 1$) it reduces to a parabolic function, but in the opposite case ($x\gg 1$) digamma function becomes a constant ($\psi(1/x+1/2)\approx \psi(1/2)=-\gamma-2{\rm ln}2\approx -1.96$) and we arrive at the logarithmic dependence:
\begin{equation}
\delta \sigma(B)=\frac{e^2}{\pi h}{\rm ln}(B)+{\rm const}
\label{eqn3}
\end{equation}
It is important to note, that the localization theory in 2D predicts a {\em universal} slope value for magnetoconductivity which is a combination of fundamental constants.
\begin{figure}[b]
\includegraphics{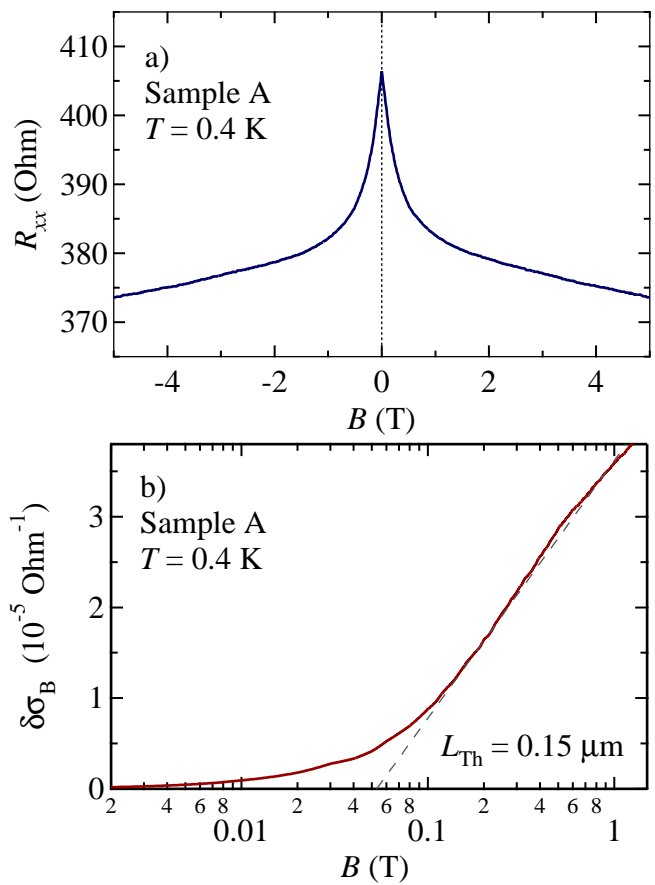}
\caption{
(a) Longitudinal resistance $R_{xx}$ as a function of $B$ showing negative magnetoresistance.
(b) Correction to conductivity due to magnetic field plotted on a semi-log scale.
Above 0.1 T we observe logarithmic rise with universal slope (c.f. dashed line), as prescribed by Eqn.~\ref{eqn2}.
}
\label{fig2}
\end{figure}

While all our samples showed similar magnetic field dependence, here we concentrate on the data acquired using sample A, which has the highest mobility, $\mu = 38$ cm$^2$/Vs ($\tau \approx$ 5.6 fs).
In Fig.~\ref{fig2}(a) we present longitudinal resistance, $R_{xx}$, as a function of magnetic field taken at $T=0.4$ K, and immediately observe sharp negative magnetoresistance peak, mainly confined to $|B|<1$ T.
To compare our observations with the theory, we convert the data to magnetoconductivity defined as $\delta \sigma(B) = \sigma_{xx}(B) - \sigma_{xx}(0)$, where $\sigma_{xx}(B)=\rho_{xx}(B)/(\rho_{xx}^2(B)+\rho_{xy}^2(B))\approx 1/\rho_{xx}(B)$.
In Fig.~\ref{fig2}(b) we plot the $\delta \sigma(B)$ in a semi-log scale and observe clean logarithmic dependence in the higher field range, i.e. at $B \ge$ 0.1 T.
Moreover, the slope value is in excellent agreement with the theory as indicated by the dotted line representing the universal slope of $e^2/\pi h$.
While the slope is universal, the intercept with $B=0$ axis is system-dependent as it bears the information about the localization properties.
Fitting our data with Eqn.~\ref{eqn2} (not shown) we estimate the value for the dephasing length $L_{{\rm Th}}\approx 0.13$ $\mu$m, which translates into the inelastic scattering time, $\tau_{{\rm in}} \approx 6.2$ ps in our sample; as expected, $\tau_{{\rm in}} \gg \tau \approx 5$ fs.

In summary, we have investigated the electrical properties of the novel, ultradense 2DES realized by P $\delta$-doping in single-crystal Si.
In particular, temperature and magnetic field dependence of the conductivity were measured down to $^3$He temperatures and magnetic fields up to 5 T.
$B$- and $T$-induced corrections to the conductivity showed characteristic logarithmic dependence, pertinent to a 2D system with weak localization effects.
Unexpected enhancement of the normalized coefficient $p$ in the correction may point onto importance of the electron-electron interaction effects.

We thank Y. C. Chang for helpful discussions on the electronic structure of this system. This work is supported by NSF-DMR9875129 (TCS), ARDA/ARO DAAD 19-00-1-0407, and DARPA-QuIST DAAD 19-01-1-0324.


\end{document}